\documentclass[lettersize,journal]{IEEEtran}
\usepackage{amsmath,amsfonts}
\usepackage{algorithmic}
\usepackage{algorithm}
\usepackage{array}
\usepackage[caption=false,font=scriptsize,labelfont=sf,textfont=sf]{subfig}
\usepackage{textcomp}
\usepackage{stfloats}
\usepackage{url}
\usepackage{verbatim}
\usepackage{graphicx}

\usepackage{cleveref}
\usepackage{multirow}

\usepackage{cite}
\begin{document}
	
	\title{Textural-Perceptual Joint Learning for No-Reference Super-Resolution Image Quality Assessment}
	
	\author{
		Yuqing Liu,
		Qi Jia,
		Shanshe Wang,
		Siwei Ma~\IEEEmembership{Senior Member,~IEEE,}
		Wen Gao~\IEEEmembership{Fellow,~IEEE,}
		\thanks{Y. Liu is with the School of Software, Dalian University of Technology,	Dalian 116620, China (e-mail:liuyuqing@mail.dlut.edu.cn).}
		\thanks{Q. Jia is with International School of Information Science and Engineering, Dalian University of Technology, Dalian 116620, China (e-mail: jiaqi@dlut.edu.cn).}
		\thanks{S. Wang, S. Ma, and W. Gao are with the School of Electronics Engineering and Computer Science, Institute of Digital Media, Peking University, Beijing 100871, China (e-mail: sswang@pku.edu.cn; swma@pku.edu.cn; wgao@pku.edu.cn)}
	}
	
	
	\markboth{Journal of \LaTeX\ Class Files,~Vol.~14, No.~8, August~2021}%
	{Shell \MakeLowercase{\textit{et al.}}: A Sample Article Using IEEEtran.cls for IEEE Journals}
	
	
	\maketitle
	
	\begin{abstract}
		Image super-resolution (SR) has been widely investigated in recent years. However, it is challenging to fairly estimate the performance of various SR methods, as the lack of reliable and accurate criteria for the perceptual quality. Existing metrics concentrate on the specific kind of degradation without distinguishing the visual sensitive areas, which have no ability to describe the diverse SR degeneration situations in both low-level textural and high-level perceptual information. In this paper, we focus on the textural and perceptual degradation of SR images, and design a dual stream network to jointly explore the textural and perceptual information for quality assessment, dubbed TPNet. By mimicking the human vision system (HVS) that pays more attention to the significant image areas, we develop the spatial attention to make the visual sensitive information more distinguishable and utilize feature normalization (F-Norm) to boost the network representation. Experimental results show the TPNet predicts the visual quality score more accurate than other methods and demonstrates better consistency with the human's perspective. The source code will be available at \url{http://github.com/yuqing-liu-dut/NRIQA_SR}
	\end{abstract}
	

	\section{Introduction}
	\IEEEPARstart{W}{ith} the rapid development of high-definition display technologies, image super-resolution (SR) has been widely investigated in advanced applications, aiming to generate high-resolution (HR) images from the given low-resolution (LR) instances. Although there are numerous image SR works in the past decades, how to estimate the quality of super-resolved images still remains challenging.
	
\begin{figure}[t]
		\captionsetup[subfloat]{labelformat=empty, justification=centering}
		\begin{center}
			\captionsetup{font=scriptsize}
			\subfloat[(a) Original~Image~\linebreak{(PSNR/SSIM/TPNet)}]{\includegraphics[width=0.45\linewidth]{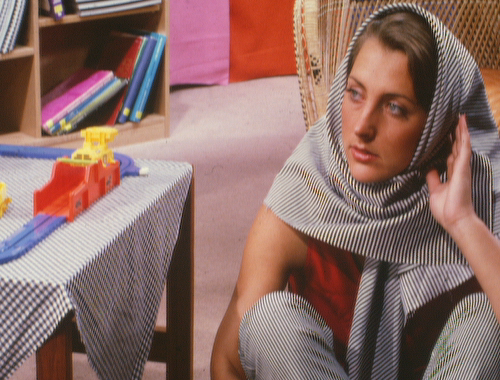}} 
			\subfloat[(b) Noise~Image~\linebreak{(18.07/0.6717/0.6330)}]{\includegraphics[width=0.45\linewidth]{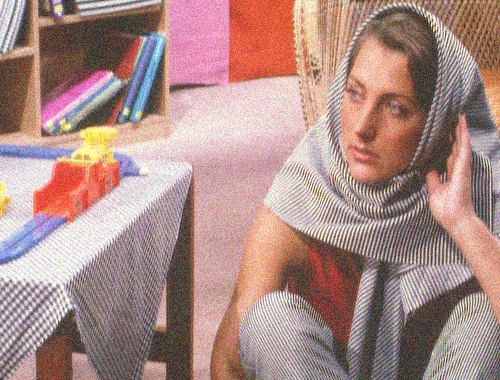}}
		
			\subfloat[(c) Blurred~Image~\linebreak{(23.01/0.6611/0.2852)}]{\includegraphics[width=0.45\linewidth]{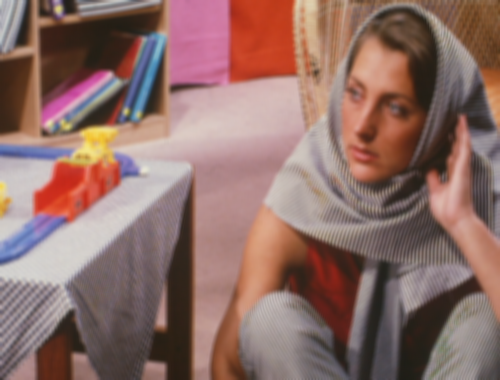}} 
			\subfloat[(d) SR~Image~\linebreak{(23.82/0.7097/0.5146)}]{\includegraphics[width=0.45\linewidth]{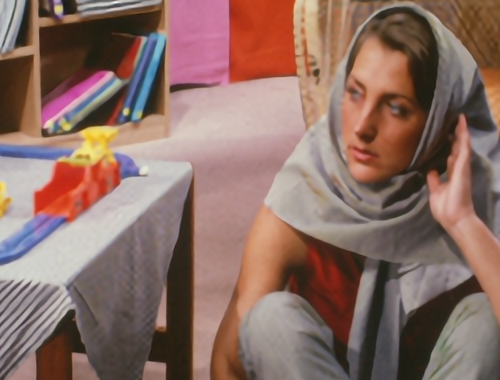}}
		\end{center}
		\caption{Comparisons among different IQA metrics. (a) original image. (b) noise image. (c) blurred image. (d) super-resolved image. The higher prediction score denotes the better quality. The proposed SR IQA method (TPNet) describes the image quality more in consistence with the visual experience.}
		\label{fig:slogan}
	\end{figure}

	There are special textural and perceptual degradation situations in the SR situation, making it hard to accurate describe the image quality. \Cref{fig:slogan} shows the comparisons among different kinds of images. We can find that there are over-smoothing areas in the SR image degrading the perceptual information, such as the edges of the arm. Besides, the generated textures in SR images are different from the original instance, as shown in the tablecloth and the scarf. General image quality assessment (IQA) metrics usually consider the simulated signal degradation and random noise, which are in low correlation with the subjective perspective of SR images. As shown in \Cref{fig:slogan}, the SR image removes the noises and blur with higher PSNR~\cite{psnr} and SSIM~\cite{ssim} scores. However, the over-smoothed and generated textures decreases the visual quality. In this point of view, SR metrics are more suitable for visual quality prediction~\cite{cviu17, qads}.
	
	Most of the recent no-reference (NR) IQA methods for SR images focus on only one kind of the degradation, which limits the representation of information loss. There are different hand-crafted descriptors to extract the textural features, such as local/global frequency features~\cite{cviu17}, spatial principal component analysis (PCA) features~\cite{zhang_icip19}, and the mean substracted contrast normalized (MSCN) coefficients~\cite{juan_access2020}. The textural features are used for evaluating the difference between the SR images and the natural ones by natural scene statistics (NSS). However, it is difficult to describe the high-level perceptual degradation by pixel-wise statistical analysis. There is also work utilizing CNN-based feature extractor to explore the perceptual information and fit the mean opinion score (MOS)~\cite{zhang_sp21}. The pre-trained deep CNN holds effective capacity for high-level perceptual feature exploration but lacks to describe the low-level textural degradation~\cite{vggloss}.
	
	The key issue of NR-IQA is to build a metric in consistence with the human vision system (HVS). According to the HVS, different areas of the image hold different importance for visual perception. However, recent NR-IQA metrics neglect to distinguish the visual sensitive information in the image and restricts the effectiveness of the prediction~\cite{cviu17, zhang_icip19, juan_access2020}. The NSS-based SR metrics take the image as a whole without considering the saliency detection. Recent CNN-based SR-IQA metrics treat different areas equally, which have no ability to highlight the visual sensitive information~\cite{qads}.
	
	In this paper, we design an end-to-end dual stream network to jointly explore the low-level textural and the high-level perceptual features from the image for NR-IQA, named as TPNet. One VGG-based~\cite{vggloss} branch is designed to explore the perceptual information, while another CNN-based branch is developed to explore the textural information. By mimicking the HVS that pays more attention to the significant areas, spatial attention is introduced to make the visual sensitive information more distinguishable. Furthermore, feature normalization (F-Norm)~\cite{isrn} is developed to boost the network representation and improve the prediction performance. Experimental results show the proposed TPNet predicts the quality more accurate than state-of-the-art IQA methods and demonstrates better consistency with human's perspective, as shown in \Cref{fig:slogan}.
	
	Our contributions can be concluded as follows:
	\begin{itemize}
		\item We design an end-to-end dual stream network named as TPNet for NR-IQA on SR images, which jointly explores the textural and perceptual information for visual perception.
		
		\item We utilize spatial attention mechanism to emphasize the significant information, resulting the visual sensitive areas more distinguishable.
		
		\item Experimental results show the proposed TPNet is more consist with the human's perspective than state-of-the-art IQA methods, rendering the accurate prediction of the subjective quality.
	\end{itemize}

	\begin{figure*}[t]
		\centering
		\includegraphics[width=0.75\linewidth]{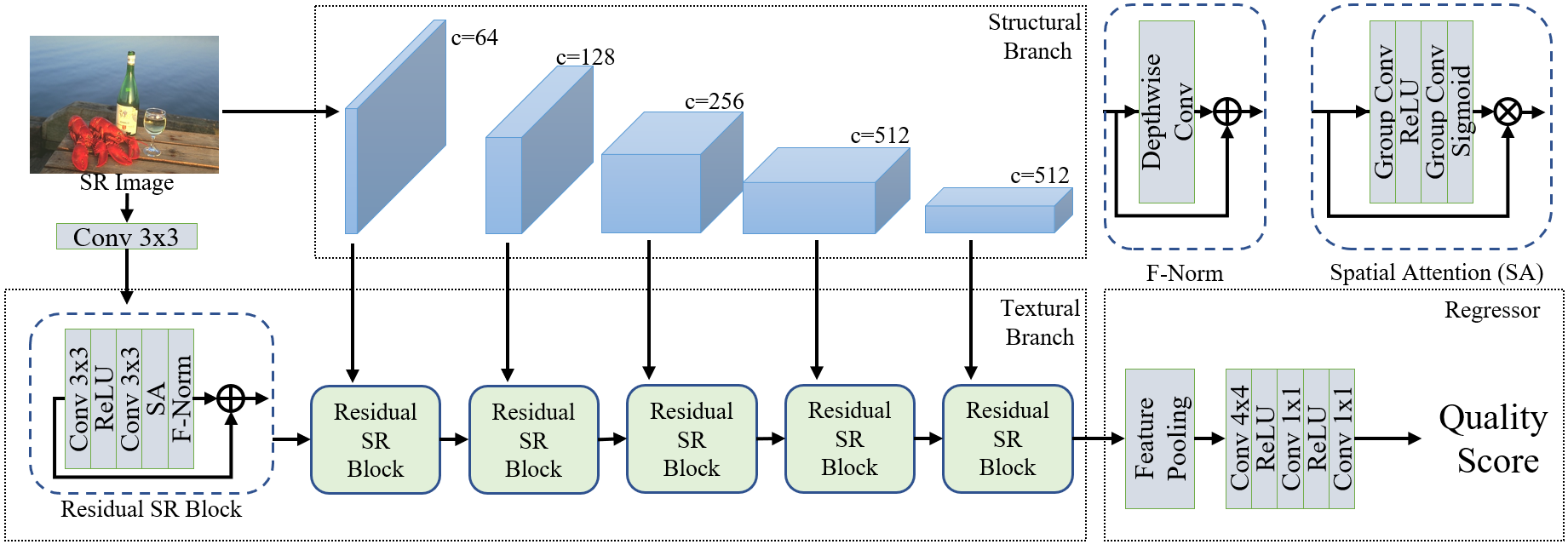}
		\caption{The architecture of TPNet. The perceptual branch utilizes a VGG-19 extractor to explore the high-level information. The textural branch stacks the proposed residual SR blocks to explore the low-level information.}
		\label{fig:net}
	\end{figure*}
	
	\section{Methodology}
	In this section, we introduce the proposed textural-perceptual joint learning network (TPNet) in the following manner. We introduce the prediction pipeline firstly. Then, we discuss the block design of the network with feature normalization (F-Norm) and the spatial attention (SA) mechanism, which are specially designed for SR features. Finally, the implementation details are described particularly.	
	
	\subsection{Prediction Pipeline}
	Given a SR image $\mathbf{I}^{SR}$, the task of NR-IQA is to predict the perceptual quality score $Q_{score}$ by a network such that
	\begin{equation}
		\label{eq:1}
		Q_{score} = TPNet(\mathbf{I}^{SR}),
	\end{equation}
	where $TPNet(\cdot)$ denotes the proposed TPNet.
	
	\Cref{fig:net} shows the design of TPNet. The network is composed of the extractor and the regressor. The extractor explores the textural and perceptual features by two dual branches. After exploration, the regressor predicts the quality score by the non-linear mapping design. There are two branches in the extractor. The perceptual branch extracts the high-level semantic information by a pretrained VGG-19 extractor~\cite{vggnet,vggloss}. Let $\mathbf{F}^P$ be the perceptual features, then there is
	\begin{equation}
		\label{eq:2}
		\{\mathbf{F}^P_i\}_{i=1}^5 = VGG(\mathbf{I}^{SR}),
	\end{equation}
	where $\mathbf{F}^P_i$ is the $i$-th perceptual feature explored by the \text{VGG-19} extractor. The channel numbers of extracted features are with $c=64, 128, 256, 512$ and $512$ separately, and the resolutions of features are halved progressively.
	
	 Correspondingly, there are stages in the textural branch to explore the low-level textural information and mix the perceptual features by the designed residual SR block. Let $\mathbf{F}^T$ be the explored textural feature, then for the $i$-th stage in the textural branch, there is
	\begin{equation}
		\label{eq:3}
		\mathbf{F}^T_{i}=RSRB([\mathbf{F}^P_{i-1}, \mathbf{F}^T_{i-1}]),
	\end{equation}
	where $RSRB(\cdot)$ is the designed residual SR block, and $[\cdot]$ denotes the channel concatenation operation. To keep the same resolution as $\mathbf{F}^P$, there is a max-pooling operation on $\mathbf{F}^T$ after each stage.
	
	 After exploration, the regressor predicts the quality score from the extracted features. There are 6 stages in the textural branch, then the quality score is predicted as
	\begin{equation}
		\label{eq:5}
		Q_{score} = Reg(\mathbf{F}^T_6),
	\end{equation}
	where $Reg(\cdot)$ is the regressor.
	
	\subsection{Residual SR Block}
	\begin{figure}[t]
		\centering
		\includegraphics[width=0.75\linewidth]{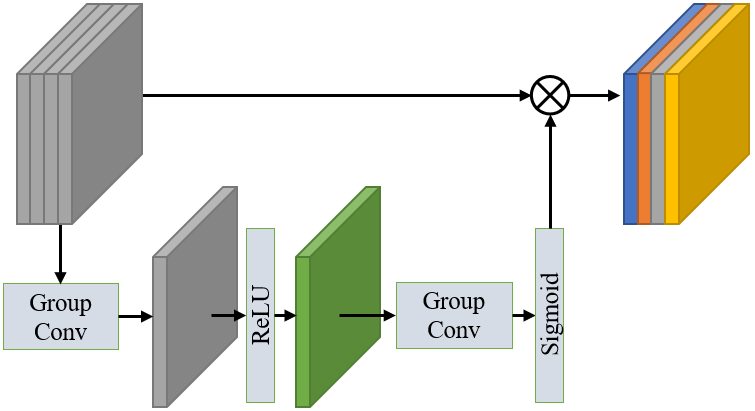}
		\caption{Design of Spatial Attention (SA).}
		\label{fig:sa}
	\end{figure}
	
	As shown in \Cref{fig:net}, the residual SR block is composed of two convolutional layers, one ReLU activation, one SA layer and one F-Norm~\cite{isrn}. The residual SR block follows the design in recent SR works and removes the batch normalization. The SA layer and the F-Norm are developed at the end of residual SR block, following the recent network designs.
	
	In the block, SA layer is utilized to make the important information more distinguishable by mimicking the human vision system (HVS), which is composed of two group convolutional layers, one ReLU activation and one Sigmoid activation. \Cref{fig:sa} shows the design of SA layer. One group convolution processes the input feature maps with group number as $c_{sa}/4$, where $c_{sa}$ is the channel number of the input feature of SA. There are $c_{sa}/4$ filters in the group convolution. After that, one ReLU activation processes the feature to introduce the non-linearity. One symmetrical group convolution restores the shape of feature with filter number as $c_{sa}$ and group number as $c_{sa}/4$. A Sigmoid activation is used to make the attention no-negative.
	
	Besides the SA layer, F-Norm~\cite{isrn} is also developed in the residual SR block to substitute the batch normalization. The upper right of \Cref{fig:net} shows the design of F-Norm. The F-Norm is composed of one depth-wise convolutional layer and one residual connection. Different from the batch normalization that widely used in different works,  F-Norm is more suitable for SR features since it can avoid the texture confusion and save the memory cost~\cite{isrn}.

	\subsection{Implementation Details}
	The perceptual branch of the TPNet is implementated by a pretrained VGG-19 network architecture. The five features are from layers with number $l=2, 7, 12, 21$, and $30$. The textural branch of the TPNet is composed of six residual SR blocks. All convolutional layers in the residual SR blocks are with filter number as $f=64$ and the kernel size as $3\times3$, except for the SA layer. 
	
	The regressor uses feature pooling to embed the features and utilizes convolutional layers to regress the quality score. Adaptive max pooling and adaptive average pooling methods compress the feature maps with size $4\times4$. Then, the convolutional layers in the regressor process the compressed features with filter number as $f=256, 64$ and $1$ separately. There is no padding in the convolutional layers, such that the regressor can generate the quality score from features with any resolution.
	
	\section{Experiment}
	\subsection{Settings}
	We choose two widely used SR-IQA datasets (CVIU-17~\cite{cviu17} and QADS~\cite{qads}) for training and testing our TPNet. CVIU-17~\cite{cviu17} proposed by Ma \textit{et al.} is one of the famous NR-IQA dataset for SR images, which contains 1620 images generated by nine traditional and CNN-based methods from six scaling factors. We randomly choose 60\% images for training, 20\% for validation and 20\% for testing. QADS is also a famous FR-IQA dataset with 980 SR images, which specially contains the results from the GAN-based method. We use the same strategy as CVIU-17 to split the dataset for training and testing. We update the TPNet for 100 epochs by Adam optimizer with learning rate as $lr=10^{-4}$. The network is implemented by the PyTorch platform, and trained on one NVIDIA GTX 3080-Ti GPU. The performances of different methods are estimated by Pearson's linear correlation coefficient~\cite{plcc} (PLCC) and Spearman's rank correlation coefficient~\cite{srcc} (SRCC). The loss function is chosen as $\ell_1$ loss between the prediction result and the mean opinion score (MOS).
	
	\subsection{Model Analysis}
	\subsubsection{Investigation on F-Norm and SA}
	\begin{table}[t]
		\centering
		\caption{PLCC/SRCC performance comparisons between F-Norm and SA on QADS dataset.}
		\label{tab:abl-fnorm}
		\begin{tabular}{|c|c|c|c|}
			\hline
			\textbf{F-Norm}& \textbf{SA}& \textbf{PLCC $\uparrow$}& \textbf{SRCC $\uparrow$} \\
			\hline
			\textbf{w/o}& \textbf{w/o}& 0.9673 & 0.9649 \\
			\textbf{w/o}& \textbf{w}  & 0.9690 & 0.9662 \\
			\textbf{w}& \textbf{w/o}  & 0.9711 & 0.9689 \\
			\textbf{w}& \textbf{w}    & \textbf{0.9720} & \textbf{0.9702} \\
			\hline
		\end{tabular}
	\end{table}

	To investigate the effectiveness of F-Norm and SA, we compare the PLCC and SRCC on the QADS dataset. \Cref{tab:abl-fnorm} shows the performance comparisons between F-Norm and SA on QADS dataset. In the table, we can find that the model with both F-Norm and SA achieves the highest PLCC and SRCC results than other methods. According to the results with and without SA (first and second lines), the SA brings 0.002 improvement on both PLCC and SRCC. From the results with and without F-Norm (first and third lines), the F-Norm leads to near 0.004 improvement on PLCC and 0.004 on SRCC. Specially, we can find from the results that F-Norm is more effective than SA with better PLCC/SRCC result. In this point of view, the F-Norm and SA boost the network performance and make the prediction more consist with the human's perspective.
	
	\begin{figure}[t]
		\captionsetup[subfloat]{labelformat=empty, justification=centering}
		\begin{center}
			\subfloat{\includegraphics[width = 0.2\linewidth]{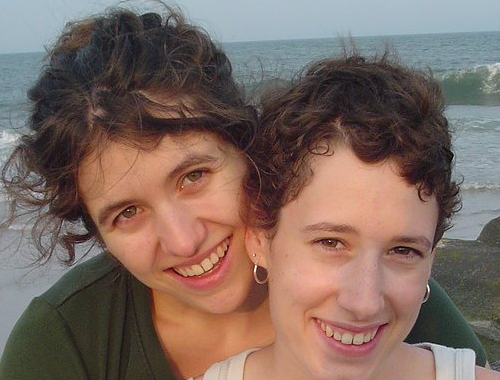}}
			\subfloat{\includegraphics[width = 0.2\linewidth]{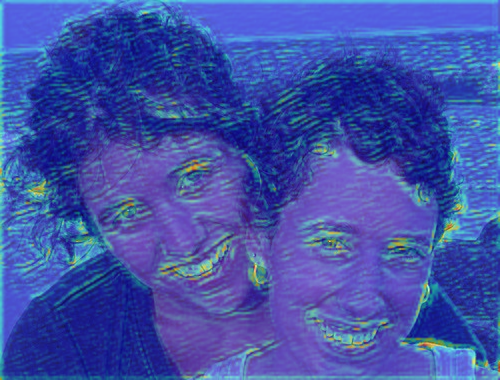}}
			\subfloat{\includegraphics[width = 0.2\linewidth]{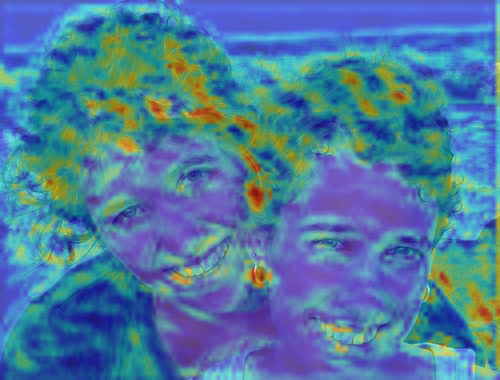}}
			\subfloat{\includegraphics[width = 0.2\linewidth]{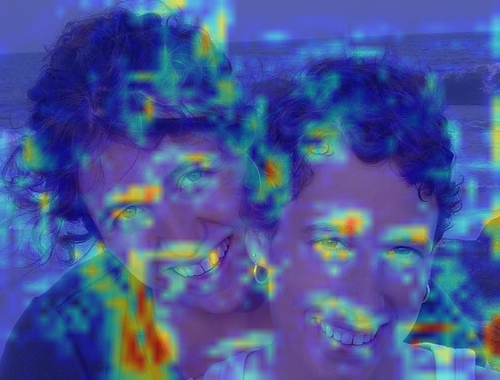}}
			\subfloat{\includegraphics[width = 0.2\linewidth]{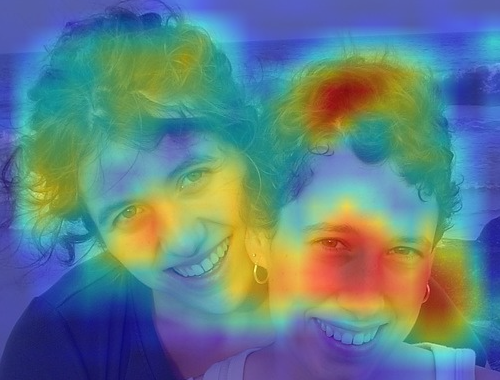}}	
			
			\subfloat[(a)]{\includegraphics[width = 0.2\linewidth]{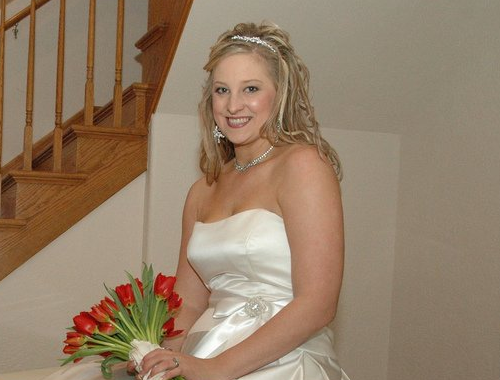}} 
			\subfloat[(b)]{\includegraphics[width = 0.2\linewidth]{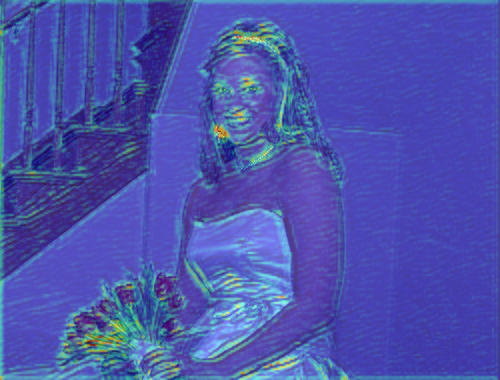}}
			\subfloat[(c)]{\includegraphics[width = 0.2\linewidth]{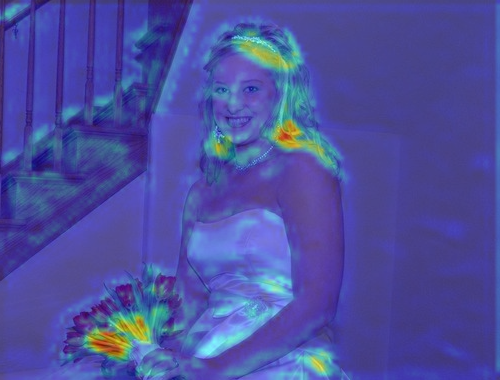}}
			\subfloat[(d)]{\includegraphics[width = 0.2\linewidth]{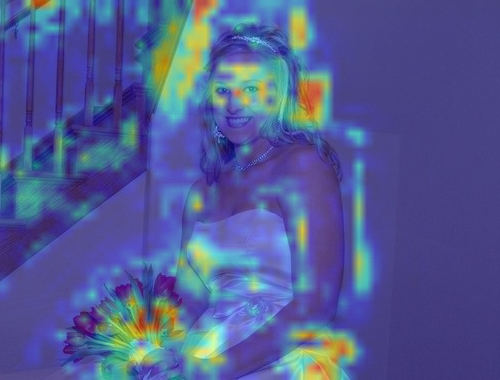}}
			\subfloat[(e)]{\includegraphics[width = 0.2\linewidth]{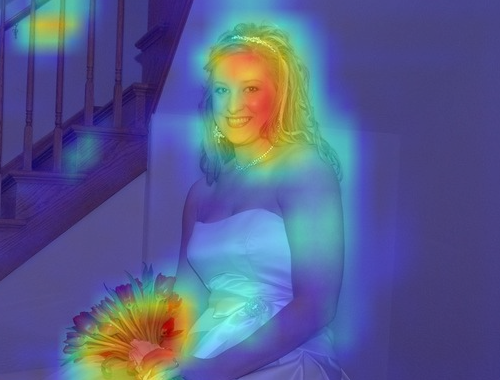}}
		\end{center}
		\caption{Visualized attention maps of different images. (a) input image. (b)-(e) the learned attention from stage $i=3$ to $6$. All of the attentions are normalized in range $0$ to $1$. The red areas mean the higher attention values, and the blue areas means the lower attention values. The visual sensitive areas become more distinguishable with the increase of stages.}
		\label{fig:abl-sa}
	\end{figure}
	
	The SA is designed to make the visual sensitive features more distinguishable. To address this point, we illustrate and demonstrate the learned attention. \Cref{fig:abl-sa} shows the visualized attention maps of different images. The (a) column denotes the original input image, and the (b)-(e) columns are the learned attention maps from stage $i=3$ to $6$, which are normalized in range $0$ to $1$. The red area means the higher value, and the blue area means the lower value. From the upper images of the figure, we can find that the complex textures become noticeable at the second stage, such as the hairs, mouths, and canthus. With the increase of stages, the attention concentrates more on the hairs and faces. We can find in the attention map of stage 6 that the eyes and haris become noticeable with a significant higher attention value. This is in accordance with the human vision system (HVS) that people usually pay attention to the faces in the picture. The similar situation can be observed in the lower images of the figure. We can find that at the stage 3, some complex textures are observed with different attention values. With the increase of stages, the face and the flower become more distinguishable, which are more sensitive to the HVS.
	
	\subsubsection{Investigation on the perceptual and textural extraction}
	\begin{table}[t]
		\centering
		\caption{PLCC/SRCC performance comparisons between perceptual and textural branches on QADS dataset.}
		\label{tab:abl-branch}
		\begin{tabular}{|c|c|c|c|}
			\hline
			\textbf{Perceptual}& \textbf{Textural}& \textbf{PLCC $\uparrow$}& \textbf{SRCC $\uparrow$} \\
			\hline
			\textbf{w/o}& \textbf{w/o}&  0.8946 & 0.8883 \\
			\textbf{w/o}& \textbf{w}  &  0.9587 & 0.9568 \\
			\textbf{w}& \textbf{w/o}  &  0.9704 & 0.9691 \\
			\textbf{w}& \textbf{w}    & \textbf{0.9720} & \textbf{0.9702} \\
			\hline
		\end{tabular}
	\end{table}
	
	In the network, we devise two branches to explore the perceptual and textural features. To show the effectiveness of the dual exploration, we compare the performances of models with different branches. \Cref{tab:abl-branch} shows the PLCC/SRCC performace comparisons between perceptual and textural branches on QADS dataset. The model only with textural branch means that no VGG-19 feature is developed. The model only with perceptual branch means that we only consider the features from VGG-19 and stacks residual SR blocks to explore the perceptual information. The model without any branch means that only one $3\times3$ convolutional layer extracts the features from the image, and the regressor predicts the score from the feature maps. In the table, we can find that the perceptual branch brings 0.08 PLCC and SRCC improvement when compared with model without any branch. Similarly, the textural branch brings 0.06 PLCC/SRCC improvement. By combing the perceptual and textural branches, the model achieves the best performance than other situations.

	\subsection{Comparison with State-of-the-Art Methods}
	We compare our model with 10 FR-IQA methods: PSNR~\cite{psnr}, SSIM~\cite{ssim}, GMSD~\cite{gmsd}, FSIM~\cite{fsim}, LPIPS~\cite{lpips}, PieAPP~\cite{pieapp}, DISTS~\cite{dists} and SFSN~\cite{zhou_qomex21}. We also compare our model with 7 NR-IQA methods: CNNIQA~\cite{cnniqa}, HyperNet~\cite{hyperiqa}, DBCNN~\cite{dbcnn}, NIQE~\cite{niqe}, NIMA~\cite{nima}, BRISQUE~\cite{brisque}, WaDIQaM~\cite{wadiqam} and NRQM~\cite{cviu17}. NIQE and BRISQUE are calculated by the MATLAB built-in function. NRQM and SFSN are calculated by the official code. We use the provided weight of NRQM for testing without further finetuning. Other implementations follow the GitHub repository\footnote{https://github.com/chaofengc/IQA-PyTorch}. For a fair comparison, we re-train the CNNIQA, HyperNet, WaDIQaM, NIMA and DBCNN under the same protocol according to our method. Specially, the images predicted by HyperNet are resized as $224\times224$ for training and testing, which follows the requirement of the model's implementation. The FR-IQA methods are not finetuned on the datasets for a fair comparison, sicne we cannot access the HR images during the no-reference assessment.
	
	\begin{table}[t]
		\centering
		\caption{PLCC/SRCC comparisons on QADS dataset among different methods}
		\label{tab:qads-result}
		\begin{tabular}{|c|c|c|c|}
			\hline
			\textbf{Type}& \textbf{Method}& \textbf{PLCC}& \textbf{SRCC} \\
			\hline
			\multirow{8}*{\textbf{Full Ref}}
			& PSNR~\cite{psnr} 			& 0.3099 & 0.3260 \\
			& SSIM~\cite{ssim} 			& 0.5187 & 0.5378 \\
			& GMSD~\cite{gmsd}			& 0.7694 & 0.7988 \\
			& FSIM~\cite{fsim}			& 0.6700 & 0.6951 \\
			& LPIPS~\cite{lpips}		& 0.6775 & 0.6782 \\
			& PieAPP~\cite{pieapp}		& 0.7481 & 0.8525 \\
			& DISTS~\cite{dists}		& 0.6739 & 0.6711 \\
			& SFSN~\cite{zhou_qomex21}			& 0.7590 & 0.8685 \\
			\hline
			\multirow{7}*{\textbf{No Ref}}			
			& NIQE~\cite{niqe}			& 0.0939 & 0.0783 \\
			& BRISQUE~\cite{brisque}	& 0.5864 & 0.6172 \\
			& CNNIQA~\cite{cnniqa}		& 0.9105 & 0.9034 \\
			& HyperNet~\cite{hyperiqa}	& 0.8413 & 0.8375 \\
			& DBCNN~\cite{dbcnn}		& 0.9477 & 0.9453 \\	
			& NRQM~\cite{cviu17}		& 0.7209 & 0.7231\\		
			& TPNet(Ours)				& \textbf{0.9720}& \textbf{0.9703} \\
			\hline			
		\end{tabular}
	\end{table}

	\Cref{tab:qads-result} shows the PLCC/SRCC comparisons on QADS dataset among different IQA methods. The FR-IQA methods are tested with the original model weights. The starred methods are re-trained on the QADS dataset. We can find that our method achieves the best PLCC/SRCC results than other works. Compared with CNNIQA, HyperNet and DBCNN that retrained under the same protocol, our method achieves 0.1 improvement on PLCC and SRCC. Specially, NRQM and SFSN are specially designed for SR assessment. NRQM is a NR-IQA metric and SFSN is a FR-IQA method. Compared with these works, our method demonstrates a significant superior performance that more consist with the human's perspective.

	In the table, we can also find that the general IQA methods usually perform no better than the SR-IQA methods. This is in accordance with our motivation that the general methods usually focus on the hand-crafted signal degradation and noise, but rarely investigate the special textural and perceptual degradation in the SR situation.

	\begin{table}[t]
		\centering
		\caption{PLCC/SRCC comparisons on CVIU-17 dataset among different methods}
		\label{tab:cviu17-result}
		\begin{tabular}{|c|c|c|c|}
			\hline
			\textbf{Type}& \textbf{Method}& \textbf{PLCC}& \textbf{SRCC} \\
			\hline
			\multirow{8}*{\textbf{Full Ref}}
			& PSNR~\cite{psnr} 			& 0.5985 & 0.5659 \\
			& SSIM~\cite{ssim} 			& 0.6322 & 0.6249 \\
			& GMSD~\cite{gmsd}			& 0.8359 & 0.8580 \\
			& FSIM~\cite{fsim}			& 0.7504 & 0.7678 \\
			& LPIPS~\cite{lpips}		& 0.8306 & 0.8220 \\
			& PieAPP~\cite{pieapp}		& 0.7841 & 0.7832 \\
			& DISTS~\cite{dists}		& 0.8642 & 0.8643 \\
			& SFSN~\cite{zhou_qomex21}	& 0.7547 & 0.8612 \\
			\hline
			\multirow{8}*{\textbf{No Ref}}			
			& NIQE~\cite{niqe}			& 0.3150 & 0.3279 \\			
			& BRISQUE~\cite{brisque}	& 0.2130 & 0.2277 \\
			& NIMA~\cite{nima}			& 0.9601 & 0.9558 \\
			& CNNIQA~\cite{cnniqa}		& 0.9280 & 0.9177 \\
			& HyperNet~\cite{hyperiqa}	& 0.8863 & 0.8836 \\
			& DBCNN~\cite{dbcnn}		& 0.9659 & 0.9602 \\			
			& WaDIQaM~\cite{wadiqam}	& 0.9254 & 0.9185 \\
			& TPNet(Ours)				& \textbf{0.9741}& \textbf{0.9720} \\
			\hline			
		\end{tabular}
	\end{table}

	Besides QADS, we also compare the performance on CVIU-17 dataset. \Cref{tab:cviu17-result} shows the PLCC/SRCC comparisons on CVIU-17 dataset. We can find that our method achieves the best performance than other works, which means the predicted scores of TPNet are more consist with the human's perspective. The starred methods are re-trained under the same protocol of our method. Compared with DBCNN and NIMA, the TPNet achieves 0.01 improvement on PLCC and SRCC. The FR-IQA methods perform better on the CVIU-17 dataset than the QADS dataset, since there are fewer CNN-based methods in CVIU-17. Even though, our method achieves much better PLCC/SRCC result than the FR-IQA methods. We do not compare TPNet with NRQM since the training data of its official model has an intersection with our testing images. Even though, our method leads to a large improvement than SFSN, which is a state-of-the-art FR-IQA method for SR images.

	\section{Conclusion}
	In this paper, we proposed a CNN-based NR-IQA method named as TPNet. Different from existing NR-IQA methods, we noticed that there are special textural and perceptual information losses in the SR situation, and devised a dual stream network for joint textural and perceptual feature exploration. Motivated by the human vision system (HVS), we developed the spatial attention mechanism to make the salient information more distinguishable and improve the accuracy of the prediction. Feature normalization (F-Norm) was also considered in the TPNet to better explore the super-resolved features. Experimental results show the proposed TPNet has achieves better performance on the QADS and CVIU-17 datasets than other state-of-the-art IQA methods.	
	
	\bibliographystyle{IEEEtran}
	\bibliography{main}

	\vfill
	
\end{document}